# Recent Progress on Excitation and Manipulation of Spin-Waves in Spin Hall Nano-Oscillators [*]

Liyuan Li(李丽媛)[1], Lina Chen(陈丽娜)[†,1,2], Ronghua Liu(刘荣华)[†,1], and Youwei Du(都有为)[1]

*1 National Laboratory of Solid State Microstructures, School of Physics and Collaborative Innovation Center of Advanced Microstructures, Nanjing University, Nanjing 210093, China*
*2 New Energy Technology Engineering Laboratory of Jiangsu Provence & School of Science, Nanjing University of Posts and Telecommunications, Nanjing 210023, China*

Spin Hall nano oscillator (SHNO), a new type spintronic nano-device, can electrically excite and control spin waves in both nanoscale magnetic metals and insulators with low damping by the spin current due to spin Hall effect and interfacial Rashba effect. Several spin-wave modes have been achieved successfully and investigated substantially in SHNOs based on dozens of different ferromagnetic/nonmagnetic (FM/NM) bilayer systems (e.g., FM = Py, [Co/Ni], Fe, CoFeB, $Y_3Fe_5O_{12}$; NM = Pt, Ta, W). Here, we will review recent progress about spin-wave excitation and experimental parameters dependent dynamics in SHNOs. The nanogap SHNOs with in-plane magnetization exhibit a nonlinear self-localized bullet soliton localized at the center of the gap between the electrodes, and a secondary high-frequency mode coexists with the primary bullet mode at higher currents. In nanogap SHNOs with out-of-plane magnetization, besides both nonlinear bullet soliton and propagating spin-wave mode achieved and controlled by varying the external magnetic field and current, the magnetic bubble skyrmion mode also can be excited at a low in-plane magnetic field. These spin-wave modes show thermal-induced mode hopping behavior at high temperature due to the coupling between modes mediated by thermal magnon-mediated scattering. Moreover, thanks to perpendicular magnetic anisotropy induced effective field, the single coherent mode also can be achieved without applying an external magnetic field. The strong nonlinear effect of spin waves makes SHNOs easy to achieve synchronization with external microwave signals or mutual synchronization between multiple oscillators with improving the coherence and power of oscillation modes significantly. Spin waves in SHNOs with an external free magnetic layer have a wide range of applications from as a nanoscale signal source of low power consumption magnonic devices to spin-based neuromorphic computing systems in the field of artificial intelligence.

[*] Project supported by the National Key Research and Development Program of China (Grant No. 2016YFA0300803), National Natural Science Foundation of China (Grant No. 11774150, No. 12074178, No. 12004171), the Applied Basic Research Programs of Science and Technology Commission Foundation of Jiangsu Province (Grant No. BK20170627), and the Open Research Fund of Jiangsu Provincial Key Laboratory for Nanotechnology
[†] Corresponding author. E-mail: linachen@nju.edu.cn
[†] Corresponding author. E-mail: rhliu@nju.edu.cn

## I. Introduction

As a novel phenomenon and emerging technique, the electrical current directly driven high-frequency oscillations of the magnetization of a nanomagnet based on magnetic multilayer structures shows an intense appeal for recent years. In 1996, Berger[1] and Slonczewski[2] firstly predicted that an electric current, which flows perpendicularly through a ferromagnetic sandwich structure of two magnetic layers connected by a nonmagnetic spacer, can transfer a spin angular momentum obtained in the first ferromagnetic layer to the second ferromagnetic layer through the spacer layer. The spin angular momentum carried by electrons acting as a spin-torque effect, also named as a spin-transfer-torque (STT), can manipulate spin dynamics of magnetic films by compensating or enhancing the intrinsic Gilbert damping of magnetic materials. STT has been intensively studied in magnetic tunnel junction (MTJ) [3] and spin-valve [4] nano-devices, which consists of a "fixed" spin-polarized magnetic layer (FM), a middle nonmagnetic layer (NM) and an active "free" FM. During an electrical current passes through the above ferromagnetic sandwich structures, the electrons are firstly polarized in the fixed FM due to the spin-dependent scattering at the Fermi surface of ferromagnetic material. Then the spin-polarized electrons extend into the second FM through the nonmagnetic space layer with a considerable spin-diffusion length. In addition to being as a high efficient spin transmission channel, the nonmagnetic spacer layer also plays a vital role to decouple the two ferromagnetic layers. As a result, the unpolarized electrical current acquires spin angular momentums inside the first fixed FM and transfers them as spin torques to the second "free" FM layer and induces magnetization switching or precession of the latter. Based on STT-driven magnetization switching, a promising memory STT-Magnetic Random Access Memory (STT-MRAM) with nonvolatile, fast, low-power, and unlimited endurance advantages is gaining significant industrial and academic attention. Meanwhile, the current-driven magnetization oscillation can generate a radio frequency (RF) microwave signal by utilizing various magnetoresistance effects (e.g., giant magnetoresistance (GMR), tunnel magnetoresistance (TMR), anisotropic magnetoresistance(AMR))[5]. Therefore, the current-driven magnetic nano-oscillators have the potential applications like a nanoscale microwave signal source of RF integrated circuits and systems for wireless communications. Besides, spin-waves, also called magnons, the collective precessional motion of spins in a magnetic system, have been proposed as a promising alternative system with low-power consumption for encoding information. The nanoscale current-driven magnetic nano-oscillators also can be used as the short-wavelength spin-wave sources of the developing magnon-based logic devices with ultralow energy consumption because they easily excite and control various coherent spin waves[6]. Finally, the magnetic nano-oscillator also processes the abundant nonlinear magnetic dynamics (e.g., mode hopping, mode transition, mode coexistence, and mutual synchronization), which make it suitable for developing ultra-fast spin-based neuromorphic computing in the field of artificial intelligence.

Very recently, researchers found another alternative approach to generate spin-torque without the accompanied charge-current occurring, named as pure spin current, in some magnetic heterostructures consisted of nonmagnetic heavy metals, topological materials and magnetic materials due to spin-momentum locking, spin Hall effect (SHE)[7] and spin Seebeck effect[8]. It is well known that the spin Hall effect has both intrinsic and extrinsic contributions, which are related to the spin-orbit coupling induced splitting of the intrinsic band structure of materials and the spin-dependence of the electron scattering on phonons and impurities, respectively[9, 10]. For utilizing the pure spin current produced by SHE, a conventional spintronic device consisting of FM/NM, in which the NM layer produces and exerts spin current to the adjacent FM layer, is structured. Except for the spin Hall effect in heavy nonmagnetic metal with the strong spin-orbit coupling[11], the interfacial Rashba-Edelstein effect[12] and the interfacial Dzyaloshinskii-Moriya interaction[13], which are related to the broken inversion symmetry of FM/NM system, are also non-negligible especially in the ultrathin FM/NM heterostructure or multilayer systems[14]. The strong interfacial Rashba-Edelstein effect also generates the additional spin-orbit torques (SOTs), including damping-like and field-like torques. Meanwhile, the interfacial Dzyaloshinskii-Moriya interaction could generate a chiral field resulting in the various nontrivial magnetic structures, e.g., spirals, skyrmions, and chiral domain walls in the thin-film system[15].

Additionally, the efficiency of SHE devices, defined as the total magnitude of angular momentum transferred to the ferromagnet during each electron passed through the device, is not limited by the magnitude of the angular momentum of an electron ($\hbar/2$). The reason is that each electron can experience multiple scattering between the ferromagnet and the SHE material during transversely pass through the spin torque devices and transfer its spin-angular momentum in each scattering event. In principle, the efficiency of spin-orbit torque generated by SHE and Rashba-Edelstein effect in the FM/NM system can exceed 100%, which is totally unlike the traditional STT-devices consisted of the MTJ and spin-valve sandwich structures. Moreover, since no electric current needs to flow through the active magnetic layer, SOT-devices can use the ultralow damping insulating magnetic materials (e.g., yttrium iron garnet, thulium iron garnet), and also will extend spintronic devices by using the enormous insulating magnetic materials and new geometries.

The purpose of this perspective review is to discuss the results of recent progress and highlight important emerging new areas of increasing interest in the studies and applications of the nonlinear dynamics of spin Hall nano-oscillators (SHNOs). This review is organized as follows. First, Section II introduces the nanoscale magnetization dynamics described phenomenologically by the Landau–Lifshitz–Gilbert equation and the dynamical characteristics of four main types of spin-waves modes driven by current-induced STT/SOT in nano-devices. Next, we discuss the complex experimental parameters-dependent nonlinear magnetization dynamics behaviors (e.g., several distinct spin-wave modes excitation[16], nonlinear coupling-induced mode hopping, mode transition, mode coexistence) in nanogap SHNOs with in-plane and out-of-plane magnetizations in Section III and IV, respectively. The characteristics of SHNOs synchronization with the external RF signal and mutual synchronization are present in Section V. The current-driven magnetic nano-oscillators with the unique nonlinear dynamic properties have wide applications in emerging new areas (e.g., novel magnon-based devices and brain-like computing chips). The two examples, as a nanoscale spin-wave source of the low-energy consumption magnon-based devices and high-speed spin-based neuromorphic computing, are discussed in Section VI. Finally, the future perspectives for the SOT-based spintronics are given in Section VII.

## II. Spin-wave excitation in magnetic nano-oscillators

### A. Description of magnetization dynamics

Similar to the spin-polarized electric currents in the conventional MTJ and spin-valve sandwich structures, the pure spin current induced by spin Hall effect can exert spin torque on the magnetization of the free layer. The spin-torque, as the Slonczewski's torque term, can be included in the Landau-Lifshitz-Gilbert (LLG) equation:

$$\dot{M} = -\gamma M \times H_{eff} + \frac{\alpha}{M_0} M \times \dot{M} + \frac{\beta}{M_0^2} M \times (M \times \hat{S}) + \tau_{FL} \qquad (1)$$

The right first term of the equation describes the conservative Larmor Precession around the effective field $H_{eff}$, which includes the applied external field $H_{app}$, the demagnetization field $H_d$, the anisotropy field $H_k$, and the current-induced Oersted field $H_{Oe}$ if a driven electrical current passes through the device. The right second term of the equation represents the intrinsic Gilbert damping term of materials during magnetization procession, and the third term describes the spin-torque (also named as the Slonczewski's torque or damping-like torque) induced by the pure spin current due to spin Hall effect, Rashba-Edelstein effect or spin Seebeck effect mentioned in the Introduction section above. Where γ is the gyromagnetic ratio, α is the Gilbert damping parameter, $\beta$ is the strength of the spin-orbit torque proportional to the ratio of spin current to charge current (also named as the spin Hall angle $\theta_{Hall} =$ $J_S/J_C$), $M_0$ is the saturation magnetization, and $\hat{S}$ is a unit vector representing the spin direction of the electron. One can find that the damping-like torque would be parallel or antiparallel to the intrinsic damping term related to the direction of charge current from Eq.(1). Therefore, the pure spin current-induced damping-like torque can completely compensate for the second damping term by adjusting the applied electric current and sustain a stable auto-oscillating magnetization around a certain axis with a specific frequency depending on the total effective field $H_{eff}$. [17]

Additionally, the pure spin current induced by the spin-orbit coupling effects also simultaneously generates a considerable field-like torque $\tau_{FL} \propto \beta M \times \hat{S}$, orthogonal with the damping torque, in some specific systems, as shown in the last term. Based on geometry symmetry of field-like torque with magnetization, however, this torque only results in a field-like dependent frequency shift of the magnetic dynamics. It may become a value comparable to the damping-like torque only in ultra-thin magnetic films with sub-nm thickness[18]. Therefore, one should note that the discussions of spin-orbit-torque driven magnetization dynamics in the FM/NM systems included in this review usually neglected the field-like torque effect.

### B. Classification of spin-waves modes

In magnetically ordered materials, a spin deviates from its equilibrium orientation axis and rotates for a sudden disturbance. Adjacent spins then follow this rotation and mobilize farther spins to rotate due to spin-spin interactions in the spin system. Finally, the disturbance propagates in the form of collective motion of spins, called spin-wave, in the principle of minimum system energy. The auto-oscillating magnetization driven by the pure spin current in an extended magnetic film can be viewed as one form of spin-waves. Compared to the uniform ferromagnet resonance (FMR) (spin-wave vector $k = 0$), there are two typical spin-torque-driven spin-waves in nano-structures with the extended free magnetic film: quasilinear propagating spin-wave and localized standing spin-wave solitons. The formation of these diverse dynamical modes depends on the various materials and device parameters, e.g., device geometry, magnitude and direction of the applied magnetic field[19], magnitude and distribution of the *DC* through device[20], magnetic anisotropy, demagnetization field or dipole field, and anisotropic chiral exchange interaction.

The linear or quasilinear propagating spin-wave, also called Slonczewski's linear mode usually has a higher oscillation frequency than $f_{FMR}$ of the uniform ferromagnet resonance, dissipates the energy from the oscillation center region through the emission of spin-waves as shown in Fig. 1(a)[21], and mostly exhibits a weak blueshift trend with increasing the excitation current based on Slonczewski's quasilinear theory studying of magnetic dynamics in nanocontact spin-torque nano-oscillators[22]. Slonczewski's quasilinear theory also successfully predicts the dispersion relation between frequency *f* and wave vector *k* of the dissipative propagating spin-wave, which is dominated by the exchange interaction. Its wave vector *k* is inversely proportional to the radius *R* of the nanocontact. These theoretical predictions have well been confirmed by microwave spectroscopy, Brillouin light-scattering techniques, and micromagnetic simulations in various conventional nanocontact spin-transfer-torque nano-oscillators (STNOs)[23] and recently discovered in SHNOs[17]. Besides, many experiments show that the orientation of the external magnetic field can control the emission direction of the propagating spin-wave, and ever it can be transformed into a localized spin-wave mode by rotating the external magnetic field from out-of-plane to in-plane of the film[24].

On the contrary to the linear propagating spin-wave mode, the nonlinear localized standing dissipative spin-wave mode, also named dissipative magnetic soliton [Fig. 1(b)[25]] (e.g., self-localized bullet mode, magnetic droplet/bubble mode), usually has a lower oscillation frequency than $f_{FMR}$ of the uniform FMR, so it does not dissipates energy by the emission of spin waves and only needs spin-torque to compensate for the dissipation of energy due to the internal damping. It is also the reason why the localized mode has a much lower critical current value $I_{th}$ than that of the linear propagating mode. As of now, several types of the localized spin-wave modes with the distinctly different spin configurations, including bullet mode, dynamical droplet mode or trivial bubble mode, topologically nontrivial skyrmion mode, and gyrotropic mode of the vortex core rotation (also called azimuthal spin waves), have been found in various spin-torque nana-oscillators with different device geometries or/and magnetic layers by numerical simulations[26] and experiments[27]. The bullet mode is usually excited in the in-plane magnetized films, which can be self-localized by the nonlinear effect of spin waves at high wave amplitudes, or localized by the local dipole-field or the effective demagnetization field. The nonlinearly self-localized bullet mode exhibits a significant redshift with increasing the excitation current due to the nonlinear frequency shift[28]. In contrast, the dipole-field localized bullet mode usually has a much less the excitation current-dependent frequency behavior, and its dynamical characteristics strongly depend on the effective potential well created by the edge or internal dipole-field[29].

Another dissipative solitonic mode, called the dynamical droplet mode, consists of a static central core with the magnetization direction opposite to the surrounding magnetic film, separated from the latter by a region where the magnetization experiences large amplitude dynamics [Fig.2(c)]. Hoefer and coworkers first theoretically or numerically proposed that the spin-wave droplet would be excited by using spin-polarized currents in the conventional multilayer nanostructures [30-32]. Then J. Akerman's group experimentally observed these exotic nonlinear excitations in nanocontact STNOs consisted of in-plane polarized layer and perpendicular magnetic anisotropy (PMA) free-layer[33]. The spatial magnetization distribution of the dynamical droplet resembles a nanoscale magnetic bubble in the magnetic system with PMA. This mode generally exhibits larger spectral intensity than the bullet mode, due to the large precession angle (core angle) of magnetization in the active dynamical region. Besides the current-induced spin torque, the droplet mode needs the magnetic anisotropy to stabilize its inverted dynamical core. Furthermore, the frequency of the droplet mode is generally expected to fall far below $f_{FMR}$ and almost independent of the excitation current, because it is nucleated by the local spin-torque and trapped in an effective dipolar potential well produced by the stray fields from the surrounding magnetic film. In contrast, the dynamics of the bullet mode is driven by the much larger direct effects of exchange interaction, and its frequency is generally only slightly below $f_{FMR}$ due to the nonlinear dipolar effects associated with large-amplitude precession[28, 34].

Besides the magnetic droplet soliton discussed above, there also exist other types of localized modes in the topologically nontrivial magnetization states, e.g., vortex and skyrmion or magnetic bubbles with a chiral domain wall. In addition to the dynamical reversed magnetization core of droplet mode, the dynamical bubble-like skyrmion [Fig.1(d) and Fig.1(e)] and vortex [Fig.1(f)] both have the chiral spin texture characterized by nontrivial topological numbers. The topological number of the vortex is 0.5, while that of the magnetic skyrmion equals 1. The dynamical bubble skyrmion is generally observed in magnetic systems with PMA and Dzyaloshinskii-Moriya interaction. Overall, these localized modes with a well-defined core exhibit some similar dynamics, e.g., the existence of energy barrier for their inversed core nucleation, the breathing, and gyrotropic core motions. So they have distinct spectral characteristics such as well-defined onset excitation current, even hysteresis behavior, sidebands around the primary frequency, which can be used to distinguish them from the localized bullet mode.

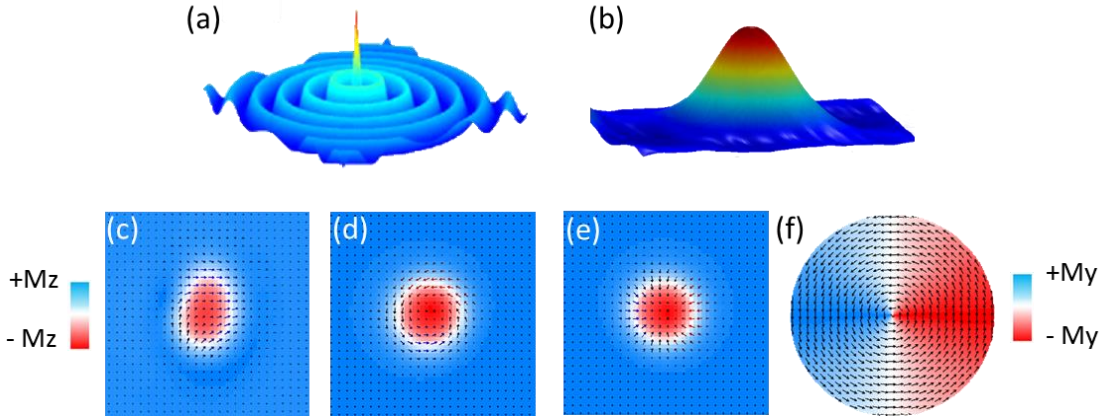

Fig. 1 Several types of magnetization dynamical modes observed in spin-torque nano-oscillators. (a)-(b) 3D spatial intensity distribution of the quasilinear propagating spin-wave mode (a) and nonlinear localized bullet mode (b), respectively. (c)-(f) Snapshots of the spatial magnetization distribution of the dynamical droplet mode or bubble mode without nontrivial topological property (c), Bloch-type (d), or Neel-type bubble skyrmion mode (e) with a topological number N = 1, and gyrotropic vortex mode (f) with a topological number N = 1/2, respectively. The color and vector represent the amplitude of the out-of-plane magnetization component Mz (left color label ) and the direction of M in (c)-(e), respectively. The right color label represents the in-plane magnetization component $M_y$ in (f). (a) and (b) are adopted from Ref. [21,25] with permission.

In the following sections, we will mainly focus on the discussion of the microwave spectral characteristics of these excitation spin-waves in spin Hall nano-oscillators. The prospective application of their nonlinear dynamics in the neuromorphic computing and magnon-based logic devices[35] also will be briefly discussed as well in the Application Section.

### III. Nanogap SHNOs with an in-plane magnetized film

**A. Mode coexistence and chaotic behavior**

In the conventional multilayer nano-structures, spin-polarized current can excite microwave magnetization oscillations of the thin free magnetic layer due to STT effect. The magnetic nano-structures, named as STNOs, generally require a relatively large current to traverse the magnetic layer, resulting in high-power and thermal degeneration or destruction of devices due to significant Joule heating. In recently developed SHE-based spintronic devices, the pure spin current generated by SHE can be used to control the magnetization of ferromagnets. In contrast to the conventional STT-based devices, the efficiency of SHE devices is not limited by the magnitude of the angular momentum of the electron. So, the latter is expected to become one type of more energy-efficient spintronic device. Moreover, since no electrical current needs to flow through the active magnetic layer, SHE devices also can utilize electrically insulating magnetic materials with low magnetic damping. Magnetization auto-oscillation driven by SHE has been recently observed by Demidov et. al.[34], using microfocus Brillouin light spectroscopy (BLS) in the Pt/Py bilayer system. The BLS experiment found that magnetization auto-oscillation is localized in a 100 nm region around the triangular nanogap electrodes, with a frequency lying significantly below its $f_{FMR}$. These features indicated that this dynamic mode is a nonlinear self-localized bullet spin-wave mode. However, the limited spectral resolution of BLS did not allow one to analyze the dynamical coherence of the oscillation in detail. Soon afterward, based on the anisotropic magnetoresistance of the Py layer that can convert its magnetization oscillation into a microwave signal, Liu and coauthors successfully achieved the electrical characterization of Py/Pt-based SHNOs via microwave spectroscopy[Fig. 2(a-b)][17, 36]. In contrast to the BLS technique, the microwave spectroscopy approach has a much high spectral resolution (<1 Hz) and also easy access to the cryogenic temperatures. Thus, the microwave spectroscopy is suitable to explore the detailed nonlinear dynamics of SHNOs, such as dynamical coherence, mode coupling, mode transition or hopping, and temperature-dependent.

Various experimental parameters (current, magnitude and angle of the magnetic field, and temperature) dependence of spin-waves excitations in nanogap SHNOs with an in-plane magnetized free layer were studied experimentally and numerically.[17, 37] The nanogap SHNOs generally consist of a microscale diameter Py/Pt bilayer disk and two triangular 100 nm thick Au electrodes. By comparing with the previous BLS experiment, the microwave spectroscopy experiments revealed a significant complexity of the dynamical behaviors in SHNOs. Liu and coauthors successfully identify several common features of the dynamical modes in these nanogap SHNOs by varying the thickness of the magnetic layer, the width of nanogap, excitation current, magnitude and angle of the magnetic field. The main panel of Fig. 2(c) and (d) are the representative generated microwave spectral of Py(5 nm)/Pt(4 nm) SHNOs[38], which reveal that a low-frequency single-mode $f_1$ is excited at small currents over a large range of experimental parameters, e.g., currents, and fields. An additional higher-frequency dynamical mode $f_2$ appears at larger currents and coexists with the primary mode $f_1$. With further current increasing, the auto-oscillation transfers to a new dynamical regime characterized by a very broad spectral feature, suggesting the emergence of multi-modes excitation. The oscillation frequencies of two modes are both below than $f_{FMR}$ of nano-device obtained by spin-torque ferromagnet resonance (ST-FMR) technique. The micromagnetic simulations further provide the additional spatial characteristics of the observed spin-wave modes. The primary auto-oscillation mode $f_1$ has an approximately elliptical spatial profile elongated in the direction of the external field $H_{tot}$ [Bottom inset in Fig.2(c)], which is well consistent with the spatial mapping results obtained in previous BLS experiments. Thus, the primary mode belongs to a nonlinear spin-wave bullet soliton localized at the center of the electrode gap. The spatial profile of the secondary mode $f_2$ has two maxima intensity offset from the active center in two opposite directions collinear with the magnetic field[Top inset in Fig.2(c)], indicating that it is also a localized mode. The simulation also demonstrated the secondary mode $f_2$ was created and stabilized by the spatially inhomogeneous dipolar field produced by the primary bullet mode. Their spatial separation facilitates the experimentally observed coexistence of the two modes. Finally, simulation results also indicated that multi-modes excitation experimentally observed at large currents is related to the chaotic dynamics of spin-waves due to the dynamic coupling between two modes. Additionally, Fig.2(d) shows that the second harmonic of primary mode $2f_1$ also can be observed in the middle fields. It has a much higher frequency than $f_{FMR}$, suggesting that it belongs to a propagating spin-wave mode. Its propagating feature also was confirmed by the simulated spatial profile [inset of Fig. 2(d)]. The propagation spin-wave achieved in SHNOs can be used as a local spin-wave source in magnon-based devices.

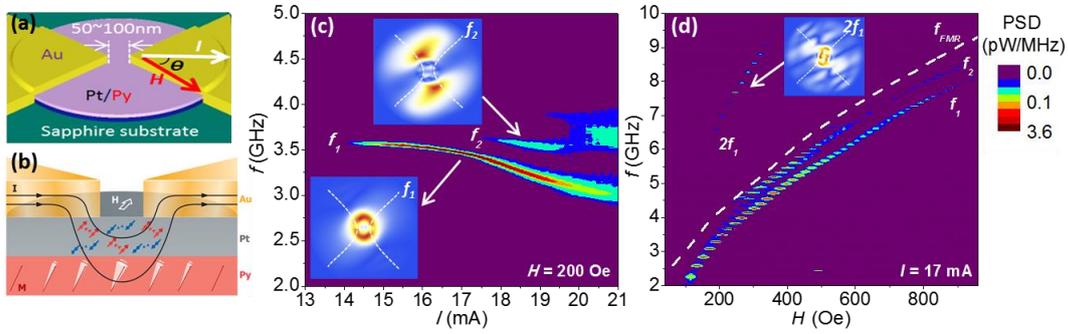

Fig. 2 (a) Schematic of the SHNO device structure and the experimental setup. (b) Schematic of the cross-sectional view of charge and spin currents distribution of SHNO device. (c) Pseudocolor map of the power spectral density (PSD) of the experimentally obtained microwave signal of nanogap SHNO based on a Py(5)/Pt(4) bilayer for varying current at $H$ = 200 Oe and $\theta$ = 60°. (d) Pseudocolor map of the experimentally obtained PSD for the varying field at $I$ = 17 mA and $\theta$ = 60°. Insets in (c) and (d): Normalized spatial maps of $m_x^2$ corresponding to the two dominant auto-oscillation modes at $f_1$ = 2.86 GHz, $f_2$ = 3.97 GHz and $2f_1$ = 5.72 GHz, respectively, which was obtained by micromagnetic simulations. Reproduced with permission from Refs. [36,38].

**B. Electric field effect of SHNOs**

According to the LLGs equation discussed in part II, the frequencies of current-induced dynamic modes significantly depend on the total effective magnetic field $H_{eff}$ including the applied external field, several magnetic anisotropies, current-induced effective field (field-like torque), and current-induced Oersted field. Several previous works have proved that the electric field can control spin-orbit coupling induced phenomena, including magnetic anisotropy and spin torques[39-42].

As the discussed SHNOs with Py/Pt bilayer above, the magnetization oscillation is excited and stabilized by SOTs. SOTs are exerted from the pure spin current generated by a combination of the SHE inside the heavy metal Pt and the interfacial Rashba effect caused by the spin-orbit coupling at the interface between Pt and Py. Therefore, it should be an attractive approach to develop spintronic devices if the current-induced spin-orbit torques and electric field effects on spin-orbit coupling related phenomena can be combined in one single device. Liu and coworkers reported a three-terminal SHNO, which can effectively combine these two effects[37], as shown in Fig. 3(a). The device also is based on a 4 um diameter Py(3 nm)/Pt(2 nm) disk with nano-gapped two Au(100 nm) tip electrodes, which is the same as above. Additionally, the controlling electric field is applied by the Cu (20 nm) gate electrode underneath Py/Pt bilayer and electrically separated each other by a 20 nm thick insulating AlOx. Besides pure spin current generated by a combination of the SHE inside Pt layer and the Rashba effect at the Pt/Py interface, several works reported that the interface between a ferromagnet and insulating oxide capping layer also could generate a considerable Rashba-like torque[43].

Firstly, the field-dependent uniform FMR mode of this three-terminal SHNO was characterized by using the ST-FMR technique, which varified that AlO$_x$/Py/Pt system has an interfacial magnetic anisotropy of $K$ = 0.27 erg/cm$^2$. Figure 3(b) shows an example of the auto-oscillation spectra acquired at $I$ = 6.7 mA, and the applied field ranging from 200 to 750 Oe at zero gating voltage $V_g$ = 0 V. The generation microwave spectral can be well fitted by the Lorenzation function, suggesting that it is related to a single coherent spin-wave. According to the experimental data under the modulation of gate voltage shown in Fig. 3(c), the center frequency of oscillation shows a linear dependence on $V_g$ with the slope of 4.6 MHz/V at $I$ = 6.2 mA. Besides, the intensity of the spectral peak also varies with the gate voltage. As LLG equation and the experimental results in Fig. 3, frequency and magnitude of oscillation not only depend on the effective magnetic field $H_{eff}$ and saturation magnetization $M_s$, but also is closely related to the spin current $\tau_{ST}$ (including direction and magnitude), which suggests the excitation current would also change the voltage-dependence of FWHM, center frequency, and power of SHNO.

To get insight into physics behind the dependence of gate-voltage modulation on current, gate-voltage-driven frequency shift can be quantitatively analyzed by the following formula:

$$df/dV_g = \frac{\partial f(I,K)}{\partial K}\frac{dK}{dV_g} + \frac{\partial f(I,K)}{\partial I}\frac{dI}{dV_g}, \qquad (2)$$

where $K$ the anisotropy coefficient, $V_g$ gate voltage, and $I$ excitation current. The first term gives the contribution of gate-voltage modulating the surface/interface anisotropy, and the second term is relative to the combination of current-induced frequency redshift due to the nonlinearity of SHNO and gate-voltage modulating spin-orbit torques efficiency. One can expect that the voltage modulating interfacial anisotropy is approximately independent of the driving current, while the second term is determined by the nonlinearity of magnetic oscillator define as $\partial f(I,K)/\partial I$. Therefore, based on the results of the current-dependence of gate-voltage modulation spectra characteristics in Fig. 3(d), Liu and coauthors qualitatively estimated the modulation rate of spin-orbit torques by gate-voltage and further separated Rashba-like torque in Py/Pt and Py/AlO$_x$ interfaces and SOT arising from the SHE in bulk Pt, respectively. Therefore, in three-terminal SHNO structure, Liu et al., not only achieved the low-power voltage-control of dynamics of the localized bullet spin-wave, but also found that the voltage modulation rate can be further magnified by current due to the nonlinearity of SHNOs. The current-dependent gating voltage modulation behavior has two distinct contributions: one comes from the E-field-dependent interfacial magnetic anisotropy, another is caused by the voltage modulation of the interfacial Rashba-like torque. The former dominates the frequency shift phenomenon of the oscillation. The latter is approximately equivalent to the nonlinear frequency redshift effect caused by driving current. Except for current-induced SOT, E-field modulation of spin-orbit coupling can be used as an additional method to control the dynamics of SHNOs for their applications in frequency mixing, synchronization, and magnon-based logic.

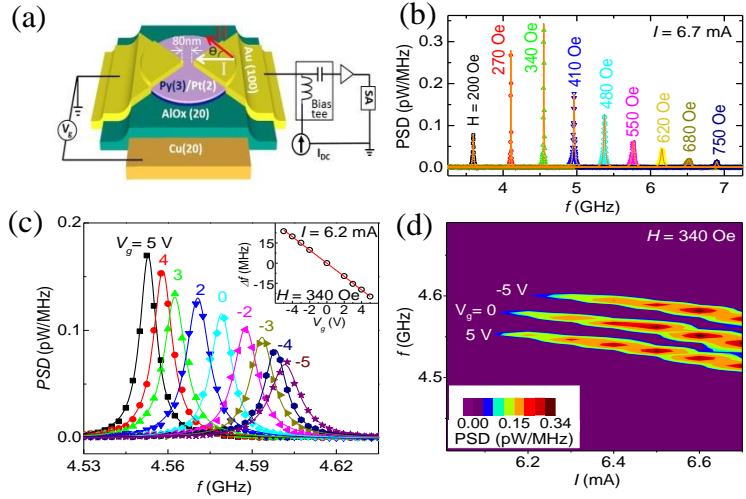

Fig. 3(a) The schematic of the SHNO structure and the experimental setup. (b) PSD signals obtained at $I$ = 6.7 mA and the labeled values of the magnetic field for the gate voltage $V_g$ = 0. (c) PSD signal detected at $I$ = 6.2 mA and the labeled values of $V_g$ varying from -5V to 5V at $H$ = 340 Oe. Inset: frequency shift *vs.* gate voltage. (d) Pseudocolor map of three-generation microwave spectral at $V_g = 0, \pm 5V$ with different currents. Reproduced with permission from Ref. [37].

## IV. Nanogap SHNOs with an out-of-plane magnetized film

### A. Propagating spin-wave and nonlinear bullet modes

As mention above, PMA can help to stabilize the magnetic bubble domain, excite propagating spin-wave mode, and localized droplet mode with an inversed core. As the conventional spin-transfer-torque nano-oscillators, one excepts that SHNOs with PMA should exhibit abundant current and magnetic field-dependent dynamics[44]. Besides a strong PMA, Pt/[Co/Ni] multilayer has a chiral interfacial

Dzyaloshinskii-Moriya interaction and a large anisotropic magnetoresistance ratio. Several previous works showed that PMA and Dzyaloshinskii-Moriya interaction could together stabilize a nontrivial magnetic bubble domain, called a topological magnetic skyrmion, in hybrid multilayer heterostructure consisted of ferromagnet and nonmagnetic heavy metal with strong spin-orbit coupling. Pt/[Co/Ni] magnetic multilayer can possess a combination effect of PMA[45], Dzyaloshinskii-Moriya interaction, and anisotropic magnetoresistance, which is considered to be the most promising system to excite magnetic bubble mode.

Liu et al. firstly designed the PMA SHNOs based on Pt/[Co/Ni] multilayer[20]. Anomalous Hall resistance loops show that the multilayer has a well-defined PMA and forms magnetic bubble states under a moderate in-plane magnetic field [Fig. 4(a)]. The angular dependent magnetoresistance exhibited a distinct 180° periodicity consistent with the anisotropic magnetoresistance of the Pt[Co/Ni] multilayer[46] [Fig. 4(b)]. After characterizing the magnetic properties of Pt/[Co/Ni] multilayer, Liu et al., performed the substantial spectroscopic measurements on SHNOs with a PMA with $H$ tilted by 5° relative to the film plane at $T$ = 140 K. At the large in-plane field, a magnetic oscillation with a frequency higher than $f_{FMR}$ is firstly excited by applying a current through the device [Fig. 4(c)], suggesting it is a propagating spin-wave mode. As the current increasing, the high-frequency mode rapidly transfers to another mode with a frequency far below $f_{FMR}$, indicating it belongs to a localized spin wave mode. The spatial intensity distributions of two spin-wave modes were given out in the insert of Fig. 4(c).

Since the static magnetization configuration of the Pt/[Co/Ni]multilayer strongly depends on the magnitude of the in-plane field, it may exhibit some distinct current-driven dynamics in the PMA SHNO at different field range[Fig. 4(d)]. Field dependent spectra obtained at $I$ = 13 mA shows that the oscillation frequency has two transitions at $H$ = 1.2 kOe and $H$~500 Oe, respectively. This latter transition is similar to the transition from the quasilinear propagating spin-wave mode to the bullet mode observed when increasing $I$ at $H$ = 1.1 kOe [Fig. 4(c)]. In contrast to the spatial coexistence of two modes observed in the Py/Pt-based SHNOs with in-plane anisotropy, one mode disappears and another mode appears precisely at the transition driven either by current [Fig. 4(c)] or by field [Fig. 4(d)]. Besides, the oscillation peak abruptly also develops sidebands around the main peak below $H = 500$ Oe. The onset field of the modulation peaks coincides with the stabilization field range of the magnetic bubble domains in the Pt/[Co/Ni] film[Fig. 4(a)]. Besides, the observed modulation sidebands spectral characteristics are reminiscent of the spin-wave droplet soliton observed in a conventional multilayer STNO with PMA of the active magnetic layer, at large fields normal to the film plane. These features suggest that the spin-wave mode should be associated with the dynamics of a nanoscale magnetic bubble trapped in the active region of SHNO. The micromagnetic simulations further confirm this interpretation and provide insight into the spatial, dynamical, and topological characteristics of three distinct modes, shown in Fig. 4(e-g). The spatial magnetization distribution of the mode observed at low fields ($H < 500$ Oe) resembles a nanoscale magnetic bubble with a chiral domain wall, also called magnetic bubble skyrmion mode. The modulation sidebands spectrum is associated with spinning or breath of the nanoscale bubble skyrmion.

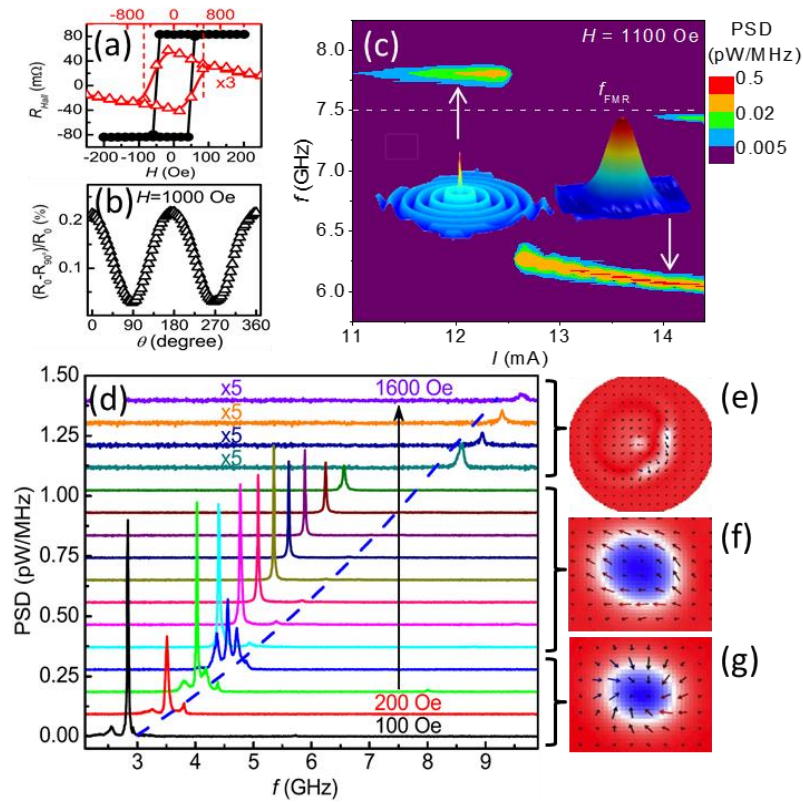

Fig. 4 (a) Anomalous Hall effect measured in a film with in-plane (triangles) and out-of-plane (circles) field at 295 K. (b) Dependence of the device resistance on the direction of in-plane field H = 1 kOe, due to the anisotropic magnetoresistance of the magnetic film. (c) Pseudocolor map of the PSD for varying current at H = 1.1 kOe. The dashed line marks the FMR of the magnetic film. Insets hint the spatial characteristics of the propagating mode (left) and bullet mode (right), respectively. (d) PSD signals detected varying field H with a step of 100 Oe at $I$ = 13 mA. (e)-(g) Snapshot of dynamical magnetization obtained from micromagnetic simulations of propagating spin-wave, bullet mode, and magnetic bubble skyrmion, respectively, at H = 300 Oe. The out-of-plane and in-plane magnetization components are represented by color and vector, respectively. Reproduced with permission from Ref. [20, 21, 25].

**B. Mode coupling-induced mode hopping**

The discussed transition between several modes above can be tuned either by current or by field. In contrast to mode coexistence facilitated by the spatial separation of different modes in SHNOs with the in-plane magnetized film[24], the different oscillation modes excited in the same spatial region are mutually exclusive in SHNOs with PMA. However, the mode hopping or periodic mode transitions commonly observed in conventional STNOs still have a chance to emerge due to the possible coupling between modes.[47, 48] The mode coexistence, hopping, and periodic transition significantly degrade their dynamical coherence and increase the linewidth. However, a nonlinear theory, developed by Slavin and coworkers, can well capture the most dynamic features of a single-mode oscillation, such as frequency redshift, driving current- and temperature-dependent linewidth.[28, 49] Recently, Iacocca et. al., developed another nonlinear theory model with consideration of the mode coupling and tried to account for these observed multimodal behaviors in experiments. The experimental verification of the mode-coupling mechanisms is essential for developing efficient microwave and spin-wave applications of SHNOs.

Recently, Chen et al. experimentally studied the temperature-dependent microwave spectra around the mode transition in SHNOs with PMA[50]. The substantial spectra characterization of SHNOs with different fields at room temperature $T$ = 295 K [Fig. 5(a-c)] reveal that the thermal fluctuation has a significant influence on spectral linewidth near mode transition region and causes mode hopping phenomenon corresponding to the appearance of a very broad linewidth. To identify the mode-coupling mechanisms in the SHNO, Chen repeated the spectroscopic measurements at a cryogenic temperature $T$ = 6 K[Fig. 5(d-f)], where thermal effects are significantly reduced. However, in contrast to the room-temperature data obtained at $H$ = 0, 700 Oe, and 900 Oe, shown in Fig. 5(a-c), the current-dependent spectra of the two dominant modes (the high-frequency propagating mode $m_P$ and the low-frequency localized "bullet" mode $m_L$) shown in Fig. 5(d-f) exhibit distinct spectral characteristics, and not noticeably affect each other near the transition between modes. These experimental results show that mode coupling causing mode hopping near the mode transition is actively suppressed, suggesting that thermal-magnon-mediated scattering rather than direct interactions between the modes is the mechanism of mode coupling in nanogap PMA-SHNO. Besides, the excellent coherent oscillation mode also can be achieved without the applied external field due to the strong PMA-induced an effective anisotropy field [Fig. 5(a) and (d)].

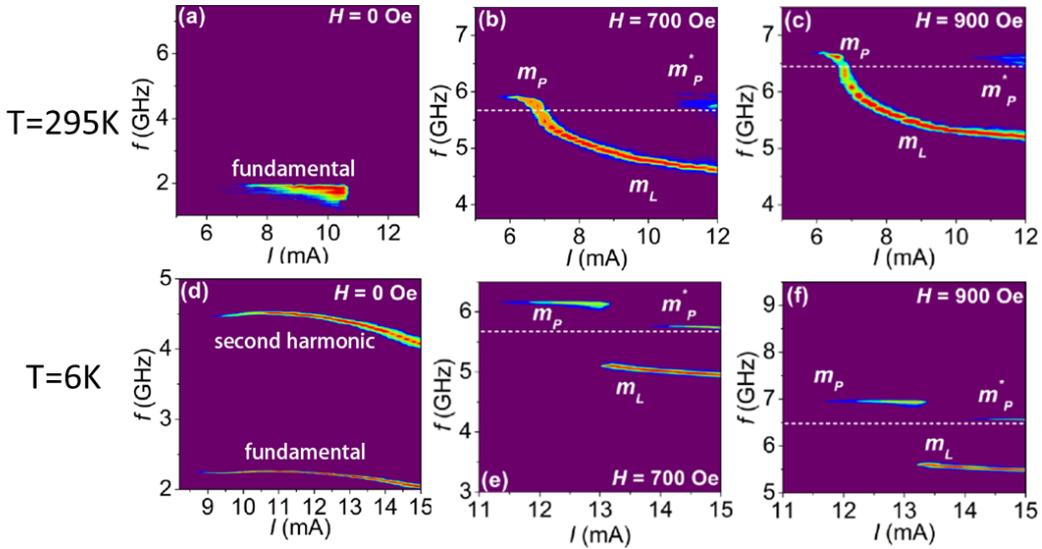

Fig. 5 Temperature effect on mode hopping. Pseudocolor maps of the dependence of the generated microwave spectra on the current at several selected fields are measured at $T$ = 295 K (a-c) and $T$ = 6 K (d-f). Reproduced with permission from Ref. [50].

## C. Magnetic droplet mode in a VNC-SHNO with oblique fields

As discussed above, the SHNO geometry, forming an in-plane nanocontact with two nonmagnetic tip electrodes, has been extensively studied to date because of its simplicity and reproducibility. However, spin current in the planar nanogap SHNO with an extended free magnetic layer prefers simultaneous excitation of two dynamical modes discussed above, degrading coherent of spin-waves at ambient temperatures due to the thermal-magnon-mediated mode hopping. Meanwhile, multimodal dynamics are also commonly observed in an alternative SHNO configuration based on an F/N nanowire due to the small frequency separation among the dynamical modes in a relatively large magnetic system. Besides, the geometry of SHNOs based on bow-tie-shaped nanoconstrictions in extended FM/NM bilayers or tapered nanowires suffers from robust mode localization by the dipolar edge fields, diminishing their thermal stability and preventing them from the exciting propagating spin-wave mode.

Recently, Chen et al., designed and experimentally demonstrated an alternative type of SHNO based on a vertical nanocontact (VNC) fabricated on a Pt/(Co/Ni) multilayer [Fig. 6(a)] and overcame these limitations of the existing SHNO geometries above[51]. In the VNC-SHNOs, the authors found that a dissipative magnetic soliton with a significant power intensity stabilized by the local injection of spin current from a single vertical nanocontact[Fig. 6(b-c)]. The droplet mode consists of an inverted core surrounded by a region experiencing large-amplitude precession, similar to the nanoscale magnetic bubble mode discussed in PMA-SHNO at low in-plane fields. In contrast to the exponential temperature dependence of the linewidth of the bullet mode in a nanogap SHNO with in-plane magnetization, the droplet mode achieved in VNC-SHNO exhibits a linear temperature dependence of the linewidth. It has higher thermal stability and a small minimum linewidth of 4.5 MHz at room temperature. The authors note that, in contrast to the bullet mode, the active dipolar potential well, created by the stray field of the inverted core, is colocalized with the precessing region in this VNC geometry, thus preventing the formation of a secondary mode coexisted with the primary bullet mode in nanogap SHNOs, which also further confirmed by the representative calculated auto-oscillation PSD spectrum and normalized spatial map of $m_x^2$ obtained by micromagnetic simulations [Fig. 6(d-e)]. This new-type VNC-SHNO provides a viable method for achieving pure spin current-driven coherent single-mode dynamical states at room temperature. This valuable VNC-SHNO geometry also can be used in low damping magnetic insulator-based SHNOs, e.g., yttrium iron garnet.

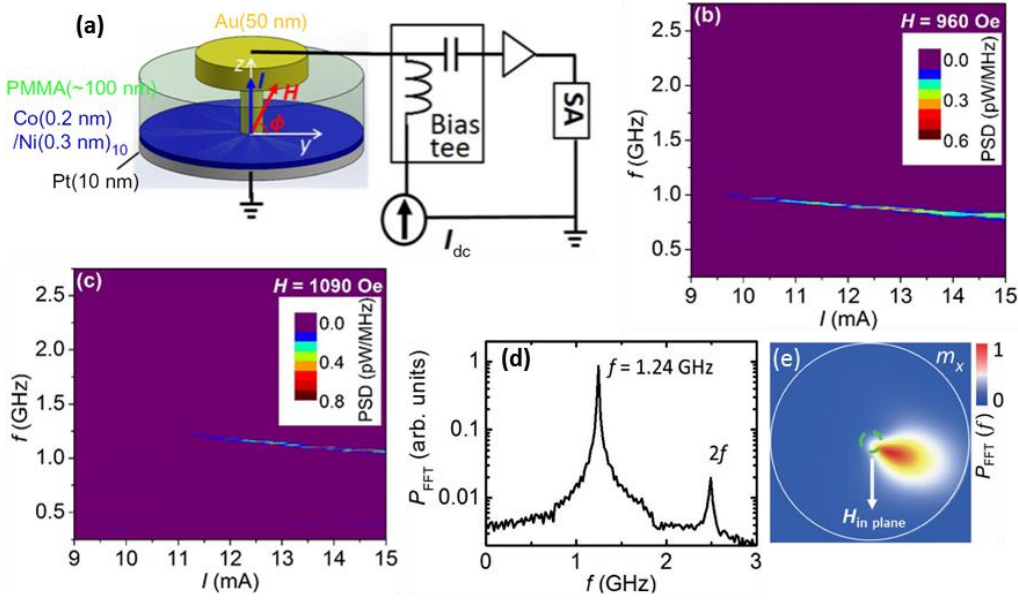

FIG. 6. Microwave spectra and micromagnetic simulation of the VNC-SHNO. (a) The device structure and the experimental setup of VNC-SHNO. (b-c) Pseudocolor plots of the current-dependent spectra of SHNO experimentally obtained at fields $H$ = 960 Oe (b), and $H$ = 1090 Oe (c) with angle $\phi$ = 82° relative to the film plane and $T$ = 295 K. (d) Representative calculated auto-oscillation spectrum at $H$ = 1000 Oe, $\phi$ = 85° and $I$ = 14 mA. (e) Normalized spatial maps of $m_x^2$ of the fundamental droplet mode. The boundary of the active simulation region and the nanocontact are marked by the large solid circle and dotted circle, respectively. Reproduced with permission from Ref [51].

## V. Synchronization of SHNOs

Based on the discussed above, the generation microwave signal quality of SHNOs for RF-applications can be enhanced by choosing suitable magnetic materials or device geometry to suppress the thermal effect and mode hopping. As a distinguishing characteristic of magnetic nano-oscillators, strong nonlinearity enables them to easily synchronize with external microwave signals[52] or other magnetic nano-oscillators with significantly different frequencies[53].

## A. Synchronization to the external RF signal

How does SHNO synchronize to the external microwave signal? The synchronization between SHNO and the external RF signals is based on the parameter excitation mechanism. The frequency of the external RF signals usually needs to approach to the fundamental frequency or high-order harmonic frequency of magnetization oscillation. Magnetic nano-oscillators can significantly improve its coherence if it achieves a phase-lock with a high-quality external RF signal.

Sakaguchi et al. used a similar SHNO with the reported by Liu et al., shown in Fig. 2(a). They found that the strong nonlinearity of SHNO makes itself easy to achieve synchronization with an external RF signal in a relatively wide frequency range. When the frequency of the external RF signal is close to twice the auto-oscillation frequency, synchronization between oscillation and the external RF signals occurs accompanying its linewidth reduces over four orders of magnitude in the synchronized regime [Fig. 7(a)][52]. SHNOs show a wide frequency range of synchronization regime, suggesting that synchronization operation using RF signals could be used as an effective method to manipulate or modulate magnetic dynamics of SHNOs for the development of novel spintronic devices.

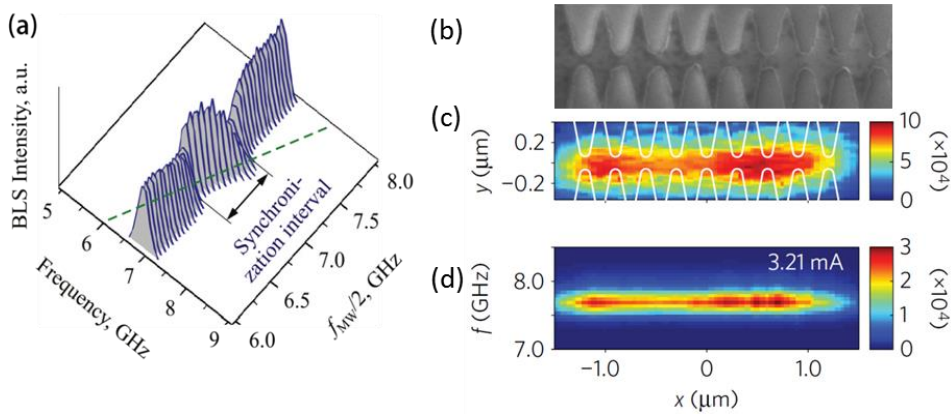

Fig. 7 (a) BLS spectra of SHNO under an external RF signal. The dashed line represents that the auto-oscillation frequency exactly follows $f_{MW}/2$. (b) A scanning electron microscope image of an SHNO array with nine 120-nm-wide nanoconstrictions each separated by 300 nm. (c-d) BLS spatial intensity map (c) and BLS frequency map (d) obtained at $I = 3.21$ mA. Reproduced with permission from Ref [52, 53].

## B. Mutual synchronization among SHNOs

Mutual synchronization between multiple oscillators results in a significant increase in the power and coherence of the generated microwave signal from the SHNOs array. By contrast to opaque and complicated STNOs based on spin-valve or magnetic tunnel junction structures, the simple planar geometry of SHNOs based on a single magnetic layer is very easier to access mutual synchronization among the SHNOs array. Rosenblum, M. G et al., reported the mutual synchronization experiments utilizing the layout of the nanoconstriction-SHNOs. The Py/Pt bilayer stripe with period multiple nanoconstrictions separated with several hundred nanometers can be regarded as multiple individual oscillators array due to the potential wells created by the dipole field[53]. Fig. 7(b) shows a scanning electron microscope image of the SHNOs stripe, which has nine 120-nm-wide nanoconstriction regions separated by distance 300 nm. The micro-focus BLS spectra directly demonstrated that all nine nanoconstriction-SHNOs could achieve mutual synchronization [Fig. 7(c-d)]. The neighboring oscillators can achieve phase-lock to each other due to the magnetic exchange interaction and dipolar coupling, and lead to mutual synchronization of all nine oscillators through their long-range self-emitted microwave current[54].

Mutual synchronization allows the enhancement of power and weak spectral coherence of magnetic nano-oscillators, which are the two essential parameters of oscillators for their applications in RF communication and microwave circuits. Meanwhile, synchronization operation also allows separately fine-tuning of their coupling constant of oscillators inside a network, which is a crucial step to mimicking basic functionalities of the human brain in nanoscale bio-inspired devices, such as vowel recognition with four coupled STNOs and temporal information processing spintronic devices discussed in detail in below part[55].

## VI. Applications of magnetic nano-oscillators

We have discussed the magnetization dynamics of several SHNOs with different magnetic materials or geometries above. SHNOs can excite several distinct spin-wave modes, including nonlinear self-localized bullet mode, quasilinear propagating spin-wave mode, magnetic bubble skyrmion mode, and magnetic droplet mode. They can be separately excited and tuned by current, magnitude and angle of the magnetic field. The multimodel behaviors (e.g., mode coexistence, mode transition, and mode hopping) were experimentally found and confirmed by micromagnetic simulation due to the mode coupling, including thermal-mediated magnon scattering, propagating spin-wave and exchange interaction or dipole field. Besides, the dynamical mode relaxation mechanism of SHNO also can upconvert low-frequency oscillation into a propagating spin wave at the higher-order harmonics of the oscillation. The propagating spin-wave can be utilized in magnon-based devices as local spin-wave sources. The nonlinear coupling between these modes also can be used for synchronization of magnetization oscillator to an external RF signal or other oscillators, implement spin-wave logical operations for analog and neuromorphic computing[56].

### A. As local spin-wave sources

Information processing in magnon-based devices or circuits is related to the three basic operations of spin-waves: generation, propagation, and detection. The coherent single-frequency spin-wave generated by spin torque nano-oscillations is spatially confined in the active magnetization oscillation region, usually preventing their applications as an efficient spin-wave source in magnonic circuits[57]. Here, Demidov and coauthors experimentally demonstrate a method to efficiently generate coherent spin-waves by nonlocal spin current injection and guide its propagation by a microscale Py waveguide[Fig. 8(a)][58]. The spin-waves are firstly excited by injecting local spin current into 5 nm thick extended Py layer in a nanocontact nano-oscillator based on Py(5nm)/Cu(20nm)/CoFe(8nm) multilayer, then are guided to the adjacent microscale waveguide made by a 20 nm thick low-damping Py strip. Fig. 8(a) shows the detailed schematic of the device structure and the experimental setup. The spatial map of BLS intensity, obtained from the micro-focus BLS spectroscopy directly, demonstrates propagating characteristics of spin-waves in Py strip waveguide from the active oscillation region [Fig. 8(b)]. Furthermore, the spatial map of the out-of-plane component $m_z$ of the dynamic magnetization (including phase and amplitude) from the simulations further confirms spin-wave propagating characteristics and well defined in the strip waveguide area, as shown in Fig. 8(c). In this structure, the spin-waves have a relatively low propagation loss with a decay length of 3 μm, sufficient large than other previously reported practical implementations of magnetic nanosystems[6, 59]. The current-induced spin torque drivens spin-waves by using nano-oscillators is more efficient than the method of using an external RF source to excite spin-waves in traditional microwave devices.

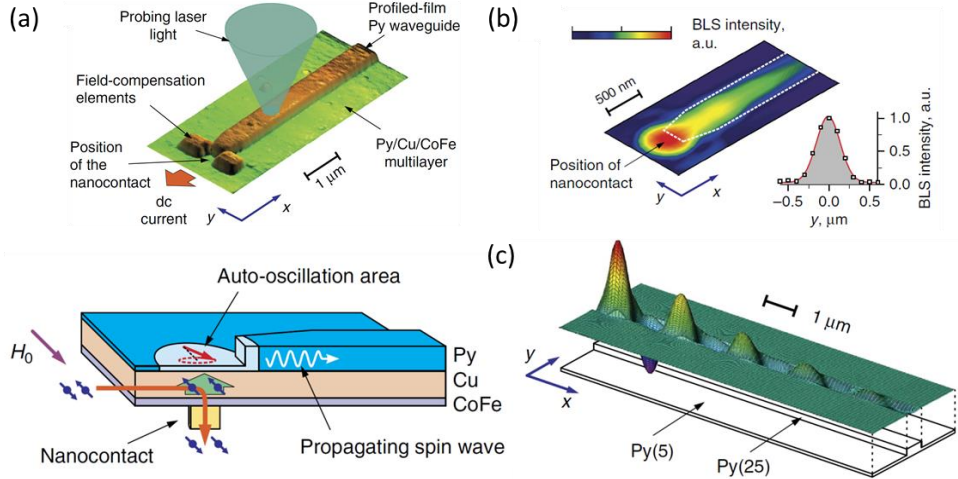

Fig. 8 Spin-waves propagation in a microscale waveguide. (a) Schematic of the device structure and the experimental setup (top panel), and AFM image (bottom panel). (b) A normalized color-coded map of the measured BLS intensity. (c) The simulation snapshot of the out-of-plane component $m_z$ of the dynamic magnetization. Reproduced with permission from Ref. [58]

## B. Spin-based neuromorphic computing

Except to generate microwave signal as an RF source and excite nanoscale spin-waves discussed above, spin-torque nano-oscillators also have a potential hung application in information processing via developing more efficient non-von

Neumann computing paradigm (e.g., analog or neuromorphic computing) by utilizing their multi-parameters controllable magnetic dynamics[60-62]. Recently, spin-torque memristors and nano-oscillators have been proposed to build artificial neural networks for applications of artificial intelligence because of their high-speed and low-power consumption[63-66]. Reservoir computing, as an important direction of the neuromorphic computing paradigm, is suitable for hardware implementations because the reservoir neural network exhibits the fixed nature of the weights between neurons. In analogy with the biological neural network [Fig. 9(a)], the physical reservoir computing network can be constructed by utilizing various physical devices with the two necessary characteristics: nonlinear and short-term memory, e.g., phase-change memories, memristors, spintronics, and optoelectronics. Fig. 9(b) shows the schematic of the experiment setup of a spin-torque nano-oscillator and its nonlinear dynamics and short-term memory[67]. STNOs and SHNOs intrinsically exhibit abundant nonlinear dynamics, including chaotic behavior, long short-term memory and fast response time (~ ns). These properties promise them as an ideal candidate for the development of high performance hardware-type artificial neural networks. Torrejon et al. experimentally achieved spoken-digit recognition with an accuracy similar to that of state-of-the-art neural networks by reservoir computing using a single vortex-type STNO[67]. Then, Liu group used the handwritten digit recognition and solving unknown nonlinear dynamic system tasks to verify the performance of the physical reservoir neural networks by micromagnetic simulations and numerically model[55]. The physical reservoir computing is built by spin-torque driven magnetic skyrmion motion memristor and vortex-type STNOs, respectively. The nonlinear dynamic system prediction task includes a second-order nonlinear dynamic system and an order 10 Nonlinear Auto-Regressive Moving Average equation (NARMA10). They exhibit the following recurrence formula respectively:

$$y_k = 0.4 * y_{k-1} + 0.4 * y_{k-1} y_{k-2} + 0.6 * u_k^3 + 0.1 \qquad (3)$$

$$y_k = 0.3 * y_{k-1} + 0.05 * y_{k-1} \sum_{i=1}^{10} y_{k-i} + 1.5 * u_{k-1} u_{k-10} + 0.1 \qquad (4)$$

Compare the theoretical values with prediction results from their spin-type reservoir neural networks after training, the authors got the normalized mean squared error (NMSE) of $1.31*10^{-3}$ and the normalized root mean square error (NRMSE) of 0.123 for the second-order[Fig. 9(c-d)] and NARMA10 nonlinear systems[Fig. 9(e-f)], respectively. These results have a better performance than the reservoir computing system with 90 metal-oxide memristors (NMSE $\simeq 3.13*10^{-3}$) in prior reports[68] as well as better than NRMSE $\simeq 0.15$ reported in a digital reservoir of 400 nodes (400 parameters needed to train).

Until very recently, Romera et al. further achieved spoken vowels recognition by training a hardware network of four STNOs according to an automatic real-time learning rule[69]. Zahedinejad et al. also achieved neuromorphic vowel recognition in two-dimensional 4*4 SHNO arrays by utilizing their synchronization and nonlinear coupling features[70]. Their experimental results demonstrate that such small spin-based hardware neural networks exhibit outstanding performance in these nontrivial pattern classification tasks. Spintronic nano-devices are highly considered as an ideal nonvolatile memory device to develop the high-performance brain-like chip for neuromorphic computing because they combine both computation and memory capabilities, and can take advantage of various physical mechanisms to implement non-von Neumann computing[71]. For example, Zeng's group found the stochastic switching behavior in an MTJ with a weak PMA free layer can be modulated by a bias voltage due to the voltage-controlled magnetic anisotropy effect, and further proved that this voltage-controlled stochastic switching could be used to perform as an adaptive neuron for building an artificial spin-neural network[72]. Neuromorphic computing can even be extended to adopt quantum algorithms to develop quantum neural networks based on the recently developed probabilistic computing[73]. We believe that how to build a large scale spin-based neural network and how to train/manipulate them efficiently for achieving high-performance neuromorphic computing would be the promising direction for the application of spintronics in the field of artificial intelligence in the coming years.

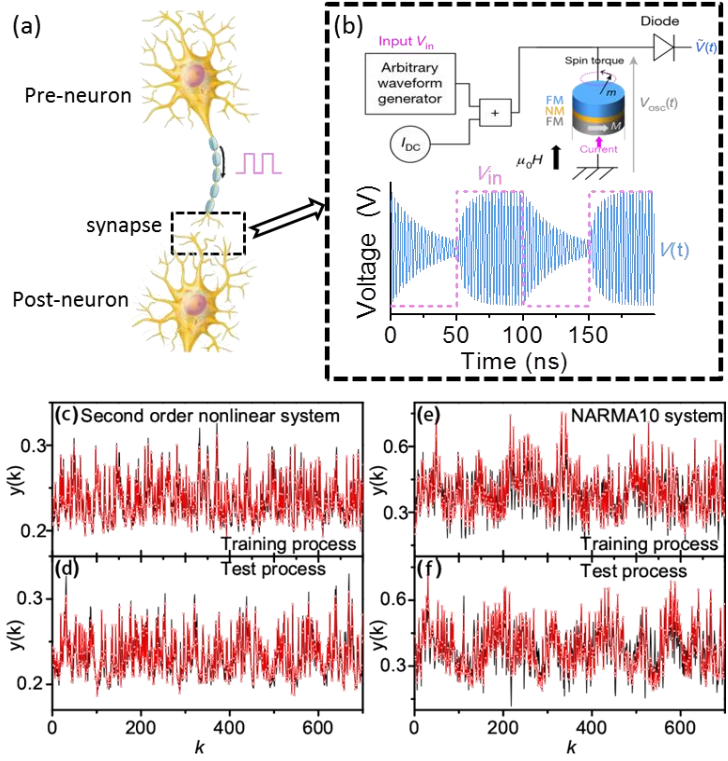

Fig. 9 (a) A schematic of the biological neural network. (b) Top: the schematic experiment setup of an STNO. Bottom: The nonlinear response (relaxation process) of the output voltage (V(t)) of SHNO with the stimulation voltage ($V_{in}$). (c-f) Training and prediction results for two nonlinear dynamic systems. (c-d) The second-order nonlinear system described by Eq. (1): theoretical output (black line) vs. prediction result (red line) in the training phase (c) and the test phase (d). (c) and (d) NARMA10 described by Eq. (2), same as (c) and (d). Reproduced with permission from Ref. [55, 67].

## VII. Conclusions：

Magnonic devices can innately avoid Joule heating induced energy loss and thermal issues in nowadays electronic nano-devices because the propagation of spin-wave is free of charge transport. Towards the aim of achieving magnonic devices, it is an essential step to find methods and materials suitable for the manipulation and detection of spin-waves. In this review, we discussed how to use the spin current arising from spin Hall effect and interfacial Rashba effect in magnetic heterostructures to electrically excite and control coherent spin-waves in both nanoscale magnetic metals and insulators. Furthermore, we systematically discussed the complex nonlinear dynamics of spin-waves of SHNOs (including the distinct spin-wave modes excitation, nonlinear coupling-induced mode hopping, mode transition, mode coexistence, and mutual synchronization) from the view of our previous microwave spectroscope experiments and micromagnetic simulations. Compared to the state-of-art sophisticated micro-focused BLS technique, the microwave spectroscope with ultrahigh-frequency resolution can easily access to various nanoscale devices as well as sorts of measurement conditions (e.g., cryogenic temperature, high magnetic field) and get more detailed features about nonlinear magnetization dynamics in nanosystems. Although extraneous effects may complicate the examination of generation spectra (e.g., spatial inhomogeneous effective fields) compared to the spatial resolution BLS, well-established techniques such as nonlinear theoretical models and numerical simulations enable us to analyze and understand magnetization dynamics from the corresponding microwave spectra, and also make such techniques powerful for both fundamental nonlinear physics studies and novel spintronic applications. SOTs, in addition to STTs, have tremendously extended the practice geometries and materials of magnetic nano-oscillators and their application fields. The dozens of different SHNOs with combining several ferromagnetic bilayer FM/NM systems and various device geometries have been investigated substantially by the optical and electrical experiments and numerical simulations. However, the details of the nonlinear processes, including nonlinear coupling and nonlinear magnetic damping, involved in spin current-induced dynamics of nano-structures have not been fully explored and understood. Further experiments and theoretical models on SHNOs will uncover new physical phenomena and effects and lead to new applications. More specifically, a promising direction is how to efficiently synchronized SHNOs array electrically by using the nonlinear interactions in new materials and geometries for achieving higher power and more coherent microwave signal generation and ultra-fast neuromorphic computing in the field of artificial intelligence.